\documentclass[10pt,letterpaper]{article}
\usepackage{opex3}
\usepackage{subfigure}
\usepackage{amsmath,amssymb}
\usepackage{bm}
\usepackage[latin1]{inputenc}
\usepackage{dcolumn}
\usepackage{graphicx}
\usepackage[amssymb]{SIunits}
\usepackage{multirow}

\begin{document}

\title{Controlling an actively-quenched single photon detector with bright light}

\author{Sebastien~Sauge,$^{1}$ Lars~Lydersen,$^{2,3}$ Andrey~Anisimov,$^{4}$ Johannes~Skaar$^{2,3}$ and Vadim~Makarov$^{2,3*}$}

\address{
$^1$School of Information and Communication Technology, Royal Institute of Technology (KTH), Electrum 229, SE-16440 Kista, Sweden\\
$^2$Department of Electronics and Telecommunications, Norwegian University of Science and Technology, NO-7491 Trondheim, Norway\\ 
$^3$University Graduate Center, NO-2027 Kjeller, Norway\\
$^4$Radiophysics Department, St.~Petersburg State Polytechnical University, Politechnicheskaya street 29, 195251 St.~Petersburg, Russia}

\email{$^*$makarov@vad1.com}



\begin{abstract*}
We control using bright light an actively-quenched avalanche single-photon detector. Actively-quenched detectors are commonly used for quantum key distribution (QKD) in the visible and near-infrared range. This study shows that these detectors are controllable by the same attack used to hack passively-quenched and gated detectors. This demonstrates the generality of our attack and its possible applicability to eavsdropping the full secret key of all QKD systems using avalanche photodiodes (APDs). Moreover, the commercial detector model we tested (PerkinElmer SPCM-AQR) exhibits two new blinding mechanisms in addition to the previously observed thermal blinding of the APD, namely: malfunctioning of the bias voltage control circuit, and overload of the DC/DC converter biasing the APD. These two new technical loopholes found just in one detector model suggest that this problem must be solved in general, by incorporating generally imperfect detectors into the security proof for QKD.\\
\end{abstract*}



\section{Introduction}
Over the past twenty years, quantum key distribution (QKD) has progressed from a tabletop demonstration to commercially available systems \cite{comqkdsystems}, with secure key exchange demonstrated up to 144 km in free-space \cite{ursin2007} and $250\,\kilo\meter$ in optical fibers \cite{stucki2009}. Security of these cryptosystems is based on the impossibility, in principle, to reliably copy an {\it a~priori} unknown quantum state, as accounted for by the no-cloning theorem \cite{wootters1982}. However, security also relies on the assumption that the optical and electro-optical devices which are part of quantum cryptosystems do not deviate from model assumptions made to establish security proofs \cite{mayers1996,gottesman2004,maroy2010,koashi2009}.

Recently, it has been demonstrated that both commercial QKD systems available on the market in 2009 could be fully cracked \cite{lydersen2010a,wiechers2011,lydersen2010b}. A tailored bright illumination was employed to remote-control gated avalanche photodiodes (APDs) used to detect single photons in these QKD systems. Note that these publications raised discussions regarding technique's applicability to QKD systems from other developers \cite{yuan2010,lydersen2010c}, as well as how such loopholes should be tackled \cite{yuan2011,lydersen2011d,yuan2011a}. In another work, a full eavesdropper has been implemented on a research system using passively-quenched APDs \cite{gerhardt2011}. The overall purpose of the work reflected in this paper is two-fold. First, we establish the generality of this attack, by extending its validity to QKD systems employing actively-quenched APDs. Second, we demonstrated two new control mechanisms in just one detector model (a commonly used commercial module, PerkinElmer SPCM-AQR \cite{spcmaqr}). The latter finding supports the opinion that efficient countermeasures must rely on a general security proof based on a sufficiently general detector model, as opposed to incremental `intuitive' technical patches.

The paper is organized as follows. In the next section, we recap the general scheme of attack which can in principle be implemented using this detector vulnerability. In sections \ref{sec:detector-control}--\ref{sec:side-effects}, we demonstrate that this particular detector, as other models tested before, fulfills the general conditions proposed for 100\% eavesdropping of the cryptographic key. We discuss countermeasures in section \ref{sec:countermeasures}, and conclude in section \ref{sec:conclusion}.

\section{Proposed attack}
\label{sec:attack-scheme}

\begin{figure}[t]
  \centering
  \includegraphics[width=133mm]{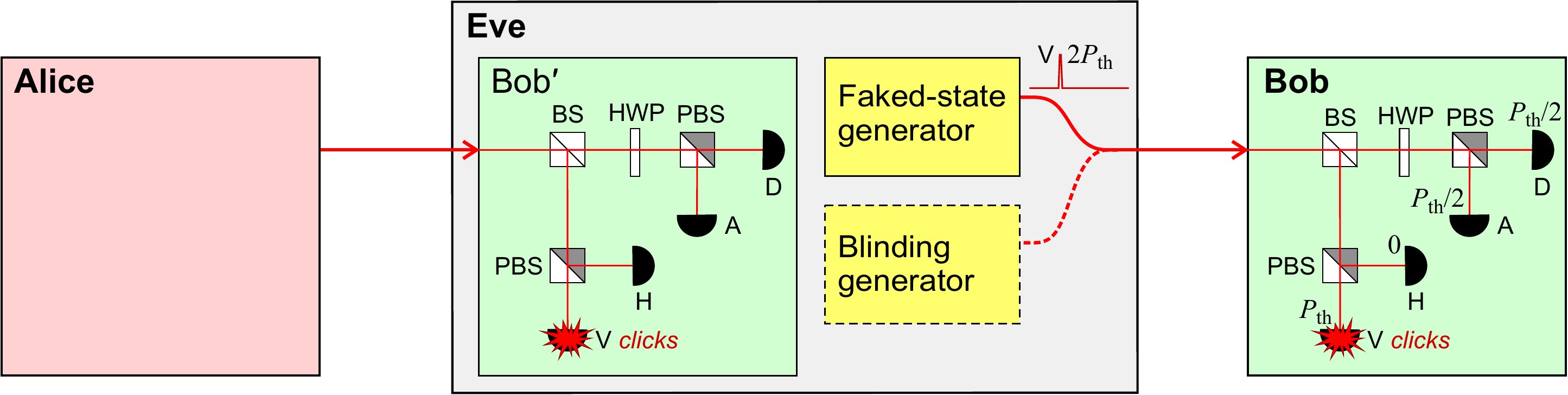}
  \caption{Intercept-resend (faked-state) attack Eve could launch against a QKD system which runs a four-state protocol with polarization coding and passive choice of basis \cite{rarity1994,peloso2009,hughes2002,erven2008}. In the example, Eve targets the detector recording vertically polarized qubits in the horizontal/vertical (H/V) basis. We assume here that detectors click controllably when illuminated by an optical pulse with peak power $\geq P_{\text{th}}$, and that they are blind (or kept blind) at power $\leq P_{\text{th}}/2$ (characteristics of the `blinding generator' potentially needed to bring detectors in this working mode will be described later). To address the target detector, Eve sends a faked state with V polarization and power $2P_{\text{th}}$, thus the V detector receives power $P_{\text{th}}$ after basis choice, and clicks. The detectors recording polarized qubits in the conjugate ($45^{\circ}$-rotated, D/A) basis each receive a pulse of power $P_{\text{th}}/2$, and thus remain blinded. In the diagram: BS, 50:50\% beamsplitter; PBS, polarizing beamsplitter; HWP, half-wave plate rotated $22.5^{\circ}$.}
  \label{fig:intercept-resend}
\end{figure}

From eavesdropper's point of view, the \emph{intercept-resend attack} provides a general framework to exploit unaccounted non-idealities or operating modes of components. In this attack, we assume that the eavesdropper Eve owns an exact replica of receiver Bob's detection apparatus, with which she intercepts and measures the state of each qubit sent by Alice. To successfully eavesdrop, Eve must resend faked states \cite{makarov2005} that will force her detection results onto Bob's in a transparent way. Ideally, the faked state should make the target detector click controllably (with unity probability and near zero time-jitter) while keeping any other detector blind (no click). In the Bennett-Brassard 1984 (BB84) \cite{bennett1984} and similar four-state protocols, Bob must detect two bit values in two bases, which can be implemented with two pairs of detectors. One pair detects bit values ``0'' and ``1'', and a second pair (not necessary with active basis choice) detects in the conjugate measurement basis, which is randomly selected prior to detection of each qubit in order to guarantee security against eavesdropping. Thus in 50\% of the cases, the qubit resent by Eve will be measured by Bob in the conjugate basis, resulting in a random outcome. Similarly, if the photonic qubit is replaced by a classical pulse of peak power $P_{\text{th}}$, an incompatible choice of basis will result in arrival of pulses of power $P_{\text{th}}/2$ at both detectors \cite{lydersen2010a, gerhardt2011}. Let us now assume that under some conditions, detectors remain blind at power $P_{\text{th}}/2$ and click controllably at threshold power $P_{\text{th}}$. With the latter pulse, Eve can selectively address the target detector without causing a click in the conjugate basis. This is illustrated in Fig.~\ref{fig:intercept-resend} in the case of a QKD system running a four-state protocol with polarization coding and passive choice of basis at Bob's side. After Bob reveals in which bit slots he has registered detections, Eve will have the same raw key bit sequence as Bob. Eve thus can extract the final secret key by listening to the classical public communication between Alice and Bob and doing the same post-processing operations as Bob \cite{lydersen2010a,gerhardt2011}. Thus, providing that the above assumption of the detector threshold behavior is satisfied, QKD systems using such detectors are vulnerable.

Let us now explain how this assumption can be fulfilled. Most QKD systems today use \emph{avalanche photodiodes} (APDs) to detect single photons \cite{cova2004}. (The two notable exceptions are continuous-variable QKD systems \cite{ralph1999,hillery2000,reid2000,heid2007,fossier2009} and those using superconducting detectors \cite{stucki2009,gol'tsman2001,verevkin2002}.) For single-photon sensitivity, APDs are operated in so-called Geiger mode, i.e., they are biased above the breakdown voltage so that an absorbed photon triggers an avalanche. (In case of gated-mode operation, the APD is biased above breakdown only during the gate time to limit noise \cite{lydersen2010a,cova2004}.) The avalanche current is sensed by a comparator before the avalanche is quenched to reset the diode. Quenching is achieved by lowering (passively or actively) the bias voltage below breakdown \cite{cova2004}. In the latter condition, however, the APD is no longer in the single-photon detection mode but behaves as a classical photodiode generating photocurrent proportional to the optical illumination. It is thus insensitive to single photons, but also to noise sources (dark counts, afterpulses). However, it is still possible to make the APD click controllably since in this classical photodiode mode, the comparator threshold translates to a classical optical power threshold $P_{\text{th}}$. Providing the threshold is well-defined, no click will ever occur at power $P_{\text{th}}/2$, and Eve has at her disposal a very general attack for breaking the security of most APD-based QKD systems.

\section{Blinding and controlling an actively-quenched single-photon detector}
\label{sec:detector-control}

In the case of the two recently-hacked commercial QKD systems operating at telecom wavelengths \cite{lydersen2010a}, transition from Geiger to classical photodiode mode was achieved by using continuous-wave (c.w.)\ bright illumination to reduce APD bias voltage below breakdown. Equivalently, raising the breakdown voltage above the fixed bias voltage by heating the APDs also led to blinding and control of the detectors \cite{lydersen2010b}.

In this paper, we illustrate further the generality of the attack by taking full control of a commercial actively-quenched detector model PerkinElmer SPCM-AQR module \cite{spcmaqr}). Until recently, the SPCM-AQR has been the only commercially available unit among actively-quenched modules. The latter account for about half of the 28 QKD experiments using non-gated detectors reported in the literature (the other half uses passively-quenched detectors) \cite{makarov2009}. It thus makes the SPCM-AQR an obvious choice for testing. Moreover, such detectors may be used after upconverting telecom-wavelength qubits into the visible and near-infrared range, where Si modules have better detection characteristics \cite{thew2009}. Free-running, actively-quenched InGaAs/InP APDs for telecom wavelenghts have also recently been introduced \cite{thew2007,id210}. Obviously, one would rather study security and control mechanisms of actively-quenched detectors before (rather than after) they get incorporated into commercial QKD systems.

\begin{figure}[b]
  \centering
  \includegraphics[width=100mm]{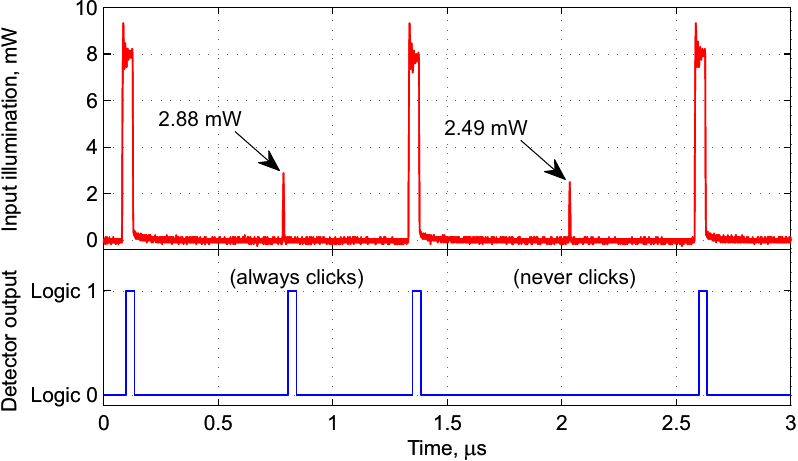}
  \caption{Oscillogram at detector output (lower trace) illuminated by bright optical pulses (upper trace) made of control pulses ($808\,\nano\meter$, $8\,\milli\watt$, $50\,\nano\second$ wide, $800\,\kilo\hertz$ repetition rate) to blind the detector, and of weaker trigger pulses ($8\,\nano\second$ wide). The trigger pulses make the detector click with unity probability and sub-nanosecond time jitter {\it only} above a certain power threshold. In the example, detector always clicks at $P_{\text{th}}=2.88\,\milli\watt$ peak power trigger pulses, never clicks at $\le$$2.49\,\milli\watt$.}
  \label{fig:spcm-aqr-control-diagrams}
\end{figure}

In the case of SPCM-AQR detector model, we achieved transition to classical photodiode mode by applying not c.w.\ (as for gated detectors) but instead bright pulsed illumination at the level of less than $10\,\milli\watt$ at $\ge$$70\,\kilo\hertz$ repetition rate. Between the pulses, the detector is blind to single photons, and does not produce dark counts or afterpulses. However, it clicks controllably if a classical light pulse $\ge$$P_{\text{th}}$ is applied, as illustrated in Fig.~\ref{fig:spcm-aqr-control-diagrams}.

\section{Blinding mechanisms}
\label{sec:blinding-mechanisms}

\begin{figure}[p]
  \centering
    \subfigure[]{\includegraphics[width=85.2mm,trim=0 -5 0 0]{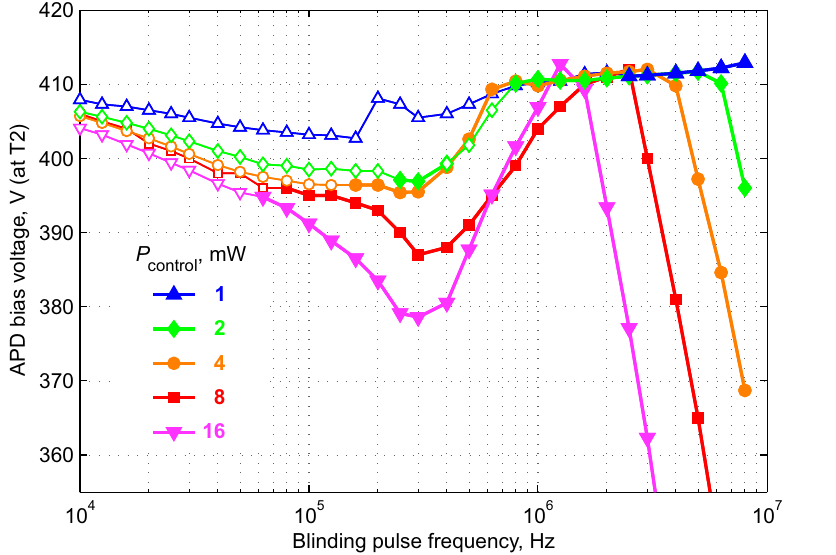}}
    \subfigure[]{\includegraphics[width=85.2mm,trim=0 -5 0 -11]{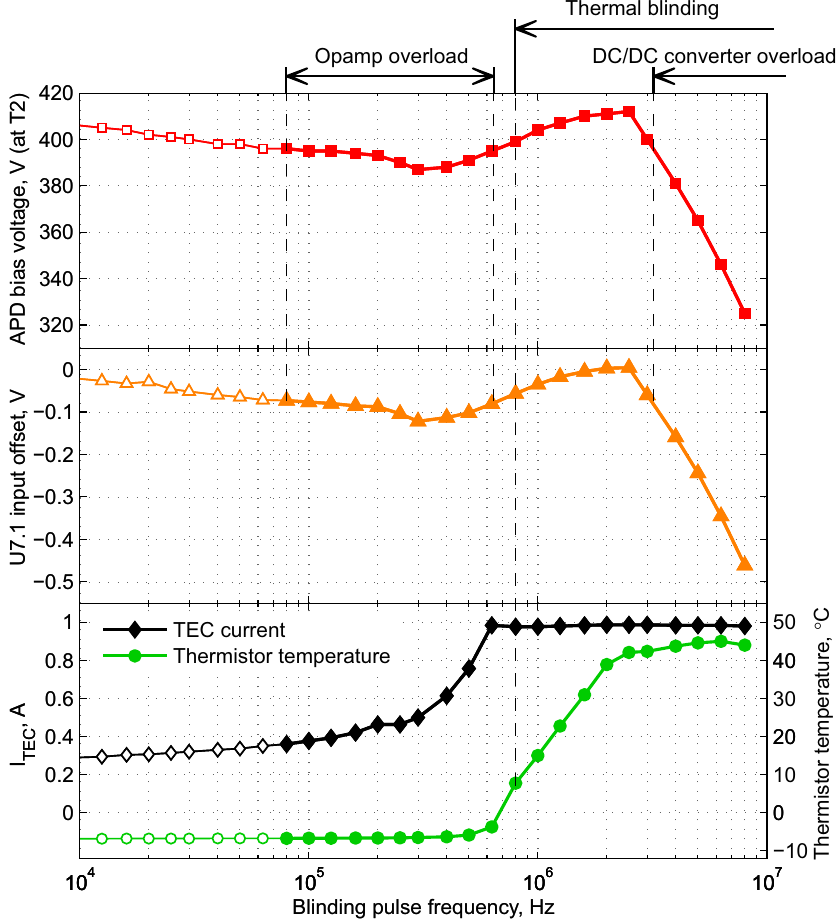}}
  \caption{Detector blinding: (a) APD bias voltage vs.\ frequency and peak optical power $P_{\text{control}}$ of rectangular $50\,\nano\second$ wide input optical pulses. Normal bias voltage at low count rate for this detector sample is $410\,\volt$ (the other detector sample we tested had bias voltage of $350\,\volt$). Filled symbols denote pulse parameters at which the detector got completely blind between the control pulses. (b) Parameters in the circuit vs.\ frequency of optical pulses with peak power $P_{\text{control}}=8\,\milli\watt$. Behavior of these parameters reveals three blinding mechanisms summarized over the top of the chart. The middle chart shows static voltage difference between the inputs of opamp, controlling the APD bias voltage (as similarity of the two top charts confirms). The lower chart shows current of the thermoelectric cooler (TEC) and the temperature of the APD as measured by a thermistor mounted nearby at the cold plate of the TEC.}
  \label{fig:spcm-aqr-attack-parameters}
\end{figure}

Fig.~\ref{fig:spcm-aqr-attack-parameters}(a) shows at which optical pulse frequencies blinding of the detector is achieved, and the corresponding bias voltage at the APD. We identified three distinct mechanisms responsible for blinding. Each mechanism is activated in a different range of control pulse frequencies, as discussed below.

\begin{figure}[t]
  \centering
  \includegraphics[width=133mm]{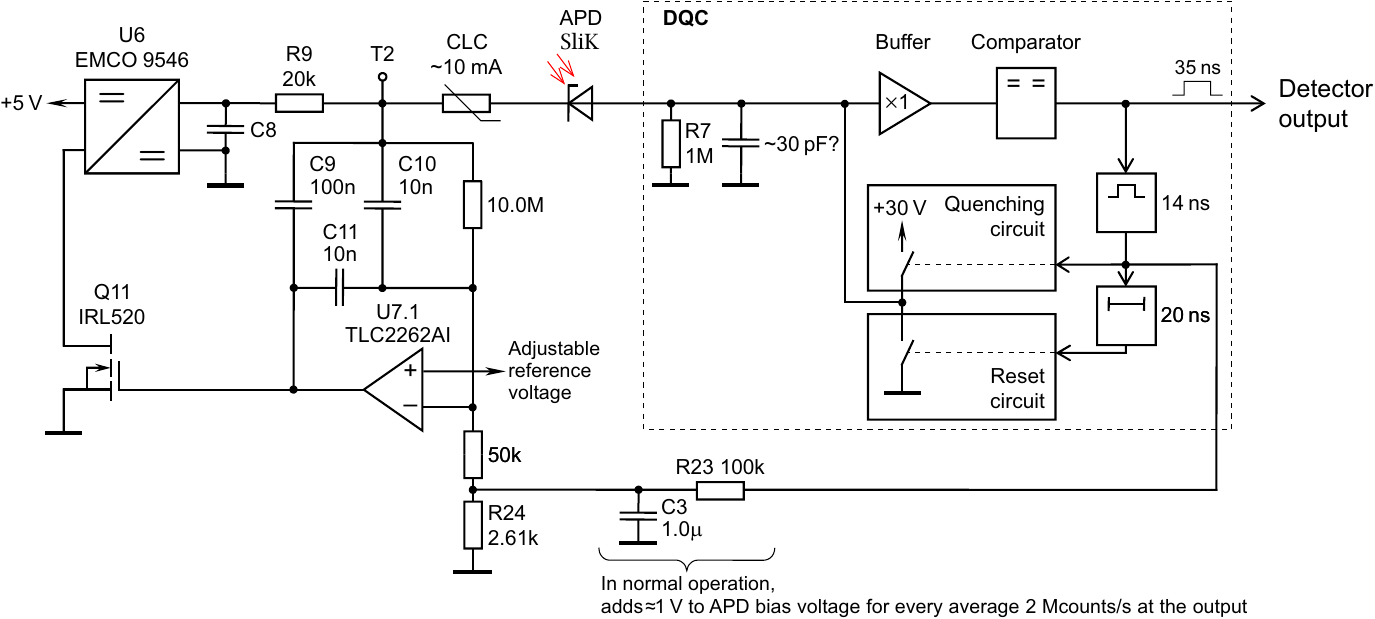}
  \caption{Simplified reverse-engineered circuit diagram of PerkinElmer SPCM-AQR module. In normal operation, the cathode of the APD (superlow-k (SliK) type \cite{dautet1993}) is biased at a constant high voltage, stabilized by a feedback loop containing an opamp U7.1 (Texas Instruments TLC2262), field-effect transistor Q11 and high-voltage DC/DC converter module U6 (EMCO custom model no.\ 9546). The anode of the APD is connected to a detection and quenching circuit (DQC). The DQC senses charge flowing through the APD during the avalanche, then briefly connects the APD anode to $+30\,\volt$ to lower the voltage across the APD below breakdown and quench the avalanche. The APD anode voltage is subsequently reset to $0\,\volt$, and the detector becomes ready for the next avalanche. (Note: the circuit diagram has been greatly simplified for the paper; do not use this figure for attempting detector repair or modification.)}
  \label{fig:spcm-aqr-detector-circuit}
\end{figure}

The first blinding mechanism corresponds to transition from Geiger to classical photodiode mode by lowering the APD bias voltage below breakdown. As the frequency of optical pulses increases, control first appears when the APD bias voltage drops by 12--15$\,\volt$ (Fig.~\ref{fig:spcm-aqr-attack-parameters}(a)). To understand why it drops, let's consider the detector electrical circuit depicted in Fig.~\ref{fig:spcm-aqr-detector-circuit}. When the APD is illuminated by a bright optical pulse, the current through it is not interrupted by the detection and quenching circuit (DQC) and is much larger than during an ordinary single-photon avalanche. A current limiting circuit (CLC) kicks in and limits the current pulse to about $10\,\milli\ampere$. This current is drawn from the capacitor C9, whose other end is connected to the output of a low-power opamp U7.1. This opamp has a specified maximum load current significantly smaller than $10\,\milli\ampere$. It gets overloaded by the current pulses, and unexpectedly develops a large static voltage offset between its inputs (see Fig.~\ref{fig:spcm-aqr-attack-parameters}(b), middle chart), which may be a behavior specific to this particular opamp integrated circuit. Yet, this negative offset effectively adds to the pre-set reference voltage at the opamp non-inverting input, and the feedback loop lowers the APD bias voltage proportionally.

\begin{figure}[t]
  \centering
  \includegraphics[width=125mm]{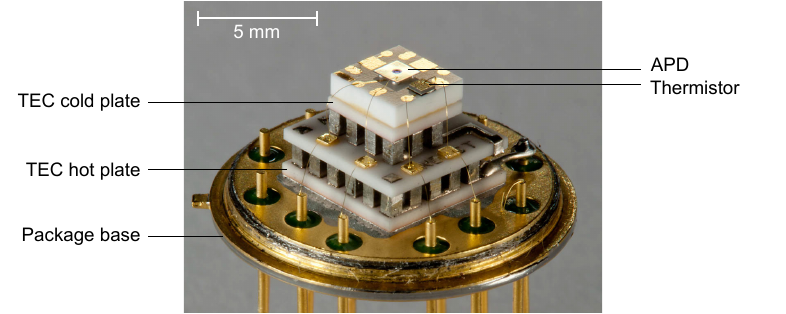}
  \caption{APD package decapsulated: the cover and fibre coupling optics have been cut off. The dark dot in the center of the APD is its photosensitive area. The APD and thermistor are mounted on the cold plate of a two-stage thermoelectric cooler (TEC). In the assembled detector, the package base is in thermal contact with an aluminum detector outer case serving as a heatsink.}
  \label{fig:apd-decapsulated}
\end{figure}

\begin{figure}[t]
  \centering
  \includegraphics[width=100mm]{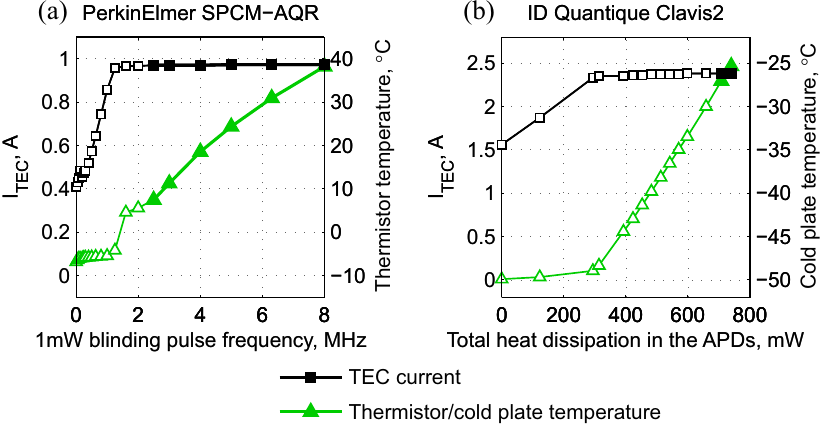}
  \caption{Comparison of thermal blinding characteristics of the PerkinElmer SPCM-AQR detector (a) to the ones reported for ID~Quantique's Clavis2 commercial QKD system \cite{lydersen2010b} (b). Filled symbols denote regime in which the detector got completely blind between the control pulses. For the SPCM-AQR, characteristics at $P_{\text{control}}=1\,\milli\watt$ are shown, because at this power thermal blinding is the only blinding mechanism.}
  \label{fig:thermal-comparison-spcm-aqr-vs-idq}
\end{figure}

At higher control pulse frequencies $\sim$$1\,\mega\hertz$, however, the disrupted opamp gets back into normal operation. At these frequencies, the duty cycle of the current pulses at the opamp output gets closer to 1/2. Since the opamp output is AC-loaded, its sourcing peak current decreases while its sinking peak current grows; they become close in magnitude and now apparently better suit opamp load capability. As a result, its large input offset disappears and the circuit raises the APD bias voltage back to the nominal value. Yet, the detector remains blind. This occurs because the APD produces more heat through electrical power dissipated in it, as the frequency of control pulses increases. In normal operation, the APD is cooled to $-7\,\celsius$ with a thermoelectric cooler (TEC), see Fig.~\ref{fig:apd-decapsulated}. The TEC heat removal capability and maximum current are inherently limited. As can be seen in the lower chart in Fig.~\ref{fig:spcm-aqr-attack-parameters}(b), after a temperature controller reaches the maximum TEC current, the APD temperature quickly rises. The raised APD temperature in turn raises its breakdown voltage (by $\approx$$1.2\,\volt/\celsius$) above the bias voltage, which also leads to blinding. The same thermal blinding behaviour has been observed before \cite{lydersen2010b}: the detectors in the commercial QKD system Clavis2 operate at a different wavelength ($1550\,\nano\meter$) and use gating instead of active quenching. Yet, as seen in Fig.~\ref{fig:thermal-comparison-spcm-aqr-vs-idq}, they have a similar response: after reaching the maximum TEC current, the APD temperature starts increasing. Eventually, after a sufficient increase in temperature, the detectors become blind. This temperature-induced increase of the breakdown voltage makes thermal blinding an attack generic in principle to all avalanche single-photon detectors.

At even higher pulse frequencies, the bias voltage drops again below breakdown, while the detector is still under control.
This is due to load capacity exhaustion of the high-voltage DC/DC converter U6 biasing the APD. 

Above, we have demonstrated three distinct blinding modes in the SPCM-AQR detector model \cite{spcmaqr}. Some QKD experiments \cite{erven2008} use a four-channel version of this detector module, PerkinElmer SPCM-AQ4C \cite{4spcmaqc}. Our preliminary analysis indicates that it has a different bias control circuit that is not susceptible to the first blinding mechanism (opamp overload). However, it is likely susceptible to both thermal blinding and DC/DC converter overload, because it uses the same APD package and the same model of DC/DC converter.

The data sheets of both detector models \cite{spcmaqr,4spcmaqc} state that exposure to intense light will reduce the count rate to zero. We could reproduce a similar effect with only one out of the two SPCM-AQR samples we tested under up to $8$--$16\,\milli\watt$ peak power, both c.w.\ and pulsed at various duty cycles and repetition rates. In that sample, a power-line monitoring circuit incorporated in the detector module (not shown in Fig.~\ref{fig:spcm-aqr-detector-circuit}) powered down the entire detector for $1$--$1.5\,\second$ following a brief time period when it was forced to count at the saturation rate of $\sim 15\,\mega\hertz$. The detector produced zero counts while it was powered down indeed, then recovered. Another peculiarity of the SPCM-AQR model is that if a weak c.w.\ illumination is present while it is being powered up (when it either recovers as above or being manually powered up by the experimenter), the DQC would latch into a `forbidden logic state' with a permanent logic high at the detector output. The detector recovers instantly from this state when the c.w.\ illumination is reduced below a certain level. (The SPCM-AQ4C circuit however seems to have no forbidden logic states.) However, we could not get controllable clicks from the detector in these blinded modes.

\section{Side effects}
\label{sec:side-effects}

There are many interesting changes that can be monitored in the circuit (Fig.~\ref{fig:spcm-aqr-attack-parameters}), however the only electrical signal an unmodified detector module currently provides to the QKD system is the detector output signal (Fig.~\ref{fig:spcm-aqr-detector-circuit}). The only side effect that betrays our attack at the detector output are clicks caused by blinding pulses with a rate of at least 70~kHz (Fig.~\ref{fig:spcm-aqr-control-diagrams}). In a general case, these clicks may increase the quantum bit error rate (QBER) observed by the legitimate parties, and a further analysis whether this attack would work is required. However, in many QKD systems Eve can arrange for these clicks to be always ignored by the legitimate parties as falling outside their post-processing gating time window. For instance, when attacking an entanglement-based system \cite{erven2008} where Eve can intercept both photons of a pair, she can arrange timing of the clicks caused by the blinding pulses to never register as coincidences between Alice and Bob. In this case, the control pulses will have zero contribution to the QBER. Similarly, if the system uses a pulsed source \cite{hughes2002, gordon2004}, the clicks can be timed to fall outside Bob's qubit time window and thus will not increase the QBER. In free-space systems operating in daylight \cite{peloso2009,hughes2002}, the clicks may be masked as a normal background count rate; note that the blinding pulses can be irregularly spaced to make them look more like background counts. We remark that the blinded state has some inertia (especially in the case of thermal blinding \cite{lydersen2010b}) that should in principle allow Eve to apply the blinding pulses in bursts interleaved with quiet periods when only the trigger pulses are applied.

Both APDs used in the present study died suddenly, after many days of extensive testing during which they worked normally and showed no advance signs of the coming failure. This may indicate that at least some of these control regimes reduce APD lifetime, although no reliable conclusions can be drawn from our limited testing. A study of failure mechanisms caused by bright light may be interesting, however it is much more challenging and expensive, and thus lies completely outside the scope of this paper. While we have exceeded the absolute maximum rating on the peak light intensity of the detector module \cite{spcmaqr}, note that Eve is never limited by manufacturer's specifications.

\section{Countermeasures}
\label{sec:countermeasures}

The major difference between public-key cryptography and quantum key distribution is that there are security proofs for the latter. However, these security proofs are based on models of the devices used in quantum key distribution. The fact that bright illumination attacks are applicable against a detector implementation, shows that this detector implementation is not within the device model of any security proof. However, the abovementioned difference from the public-key cryptography also requires loopholes to be closed differently than before. Rather than just avoid the specific attack, one must alter the implementation and/or the security proofs to re-ensure that the devices are within the models of the security proofs. Otherwise, the quantum key distribution implementation is no longer provably secure, and thus has no advantage over key distribution based on classical cryptography.

Countermeasures have been extensively discussed for gated APD-based detectors \cite{lydersen2010a,lydersen2010b,yuan2010,lydersen2010c,lydersen2011a,yuan2011,lydersen2011d,yuan2011a}. That discussion shows clearly that avoiding specific attacks does not necessarily re-establish security. The most frequently proposed countermeasure to prevent blinding consists of using an optical power meter (or the APD itself as an optical power meter \cite{yuan2010,yuan2011,yuan2011a}) with a classical threshold at Bob's entrance \cite{lydersen2010a,yuan2010,lydersen2010c,yuan2011,yuan2011a}. For such a countermeasure, the classical threshold for the optical power must originate from a more general security proof \cite{lydersen2010a}, otherwise the provable security is not re-established. In fact, a recent study \cite{lydersen2011b} has shown that for gated APD-based detectors, control can be achieved with faint trigger pulses containing less than 120 photons per pulse, already making the attack very difficult to detect with an optical power meter. For gated detectors, the so-called ``bit-mapped gating'' scheme \cite{lydersen2011a} seems to make the implementation compatible with certain security proofs \cite{maroy2010}, and may thus re-establish security. One possible way of further development down this path would be to test single-photon sensitivity of Bob's APDs at random times by a calibrated light source placed inside Bob \cite{lydersen2011a}. The results from this testing may then be used as an input to a security proof that incorporates detector deficiencies and non-linearities \cite{maroy2010}. Alternatively, countermeasures for all detectors considered may also include monitoring the APD bias voltage, current and temperature \cite{lydersen2010b}, with the efficiency of each countermeasure to be assessed in the framework of a general security proof. Although development of countermeasures has begun \cite{lydersen2010a,lydersen2011a,yuan2010,lo2011,braunstein2011}, no definite countermeasure has been finalized and tested by hacking at this time \cite{lydersen2010c}. 

If one simply wishes to avoid the specific bright-illumination attacks without re-establishing provable security, there are numerous options. One could replace the APD with a beamsplitter distributing photons to two APDs, and to look for coincidences as a signature of eavesdropping. One could also monitor any of the parameters presented in Fig.~\ref{fig:spcm-aqr-attack-parameters}, or any other parameter such as the background detection rate, revealing detector control attempts. However we don't see why anybody would then prefer this QKD system that is not provably secure, over cheaper and much more convenient classical cryptography systems.

\section{Conclusion}
\label{sec:conclusion}

In view of this study, complemented by the ones made on other APD models \cite{lydersen2010a,wiechers2011,lydersen2010b,gerhardt2011}, we estimate that most of the QKD systems existing today are potentially vulnerable to our attack. The only `detector-dependent' aspect here is the type of bright illumination (none, c.w.,\ or pulsed) required to bring a particular APD into the classical photodiode regime. We remark that very similar bright-light control method is also applicable to a superconducting nanowire based single photon detector \cite{lydersen2011c}. While we have not analysed actively-quenched detectors of manufacturers other than PerkinElmer, the above experience suggests that there is a chance they may be vulnerable to some of the already studied, as well as new yet-to-be-found blinding and control mechanisms.

While bright-light detector control is, strictly speaking, not equivalent to the hack of a QKD system, we note that for one of the APD models, a full eavesdropper based on bright-light detector control has previously been implemented and tested under realistic conditions on a $290\,\meter$ experimental entanglement-based QKD system \cite{gerhardt2011}. In view of this demonstration, we consider that closing the exposed loopholes in a provable way should take precedence over confirming that a full eavesdropper can be built for all APD models.

Our work emphasizes the need to investigate thoroughly vulnerabilities originating from unaccounted physical non-idealities of QKD components.

\section*{Acknowledgements}
Financial support is acknowledged from the Research Council of Norway (grant no.\ 180439/V30), the ECOC'2004 foundation and the Research Council of Sweden (grant no.\ 621-2007-4647).


\begin{thebibliography}{10}
\newcommand{\enquote}[1]{``#1''}

\bibitem{comqkdsystems}
Commercial QKD systems are available from at least two companies: ID~Quantique
  (Switzerland), \url{http://www.idquantique.com}; MagiQ Technologies (USA),
  \url{http://www.magiqtech.com}.

\bibitem{ursin2007}
R.~Ursin, F.~Tiefenbacher, T.~Schmitt-Manderbach, H.~Weier, T.~Scheidl,
  M.~Lindenthal, B.~Blauensteiner, T.~Jennewein, J.~Perdigues, P.~Trojek,
  B.~{\" O}mer, M.~F{\" u}rst, M.~Meyenburg, J.~Rarity, Z.~Sodnik, C.~Barbieri,
  H.~Weinfurter, and A.~Zeilinger, \enquote{Entanglement-based quantum
  communication over 144 km,} Nat. Phys. \textbf{3}, 481--486 (2007).

\bibitem{stucki2009}
D.~Stucki, N.~Walenta, F.~Vannel, R.~T. Thew, N.~Gisin, H.~Zbinden, S.~Gray,
  C.~R. Towery, and S.~Ten, \enquote{High rate, long-distance quantum key
  distribution over 250 km of ultra low loss fibres,} New J. Phys. \textbf{11},
  075003 (2009).

\bibitem{wootters1982}
W.~K. Wootters and W.~H. Zurek, \enquote{A single quantum cannot be cloned,}
  Nature \textbf{299}, 802--803 (1982).

\bibitem{mayers1996}
D.~Mayers, \enquote{Advances in cryptology,} in \enquote{Proceedings of
  Crypto'96,} vol. 1109, N.~Koblitz, ed. (Springer, New York, 1996), vol.
  1109, pp. 343--357.

\bibitem{gottesman2004}
D.~Gottesman, H.-K. Lo, N.~L{\" u}tkenhaus, and J.~Preskill, \enquote{Security
  of quantum key distribution with imperfect devices,} Quant. Inf. Comp.
  \textbf{4}, 325--360 (2004).

\bibitem{maroy2010}
{\O}.~Mar{\o}y, L.~Lydersen, and J.~Skaar, \enquote{Security of quantum key
  distribution with arbitrary individual imperfections,} Phys. Rev. A
  \textbf{82}, 032337 (2010).

\bibitem{koashi2009}
M.~Koashi, \enquote{Simple security proof of quantum key distribution based on
  complementarity,} New J. Phys. \textbf{11}, 045018 (2009).

\bibitem{lydersen2010a}
L.~Lydersen, C.~Wiechers, C.~Wittmann, D.~Elser, J.~Skaar, and V.~Makarov,
  \enquote{Hacking commercial quantum cryptography systems by tailored bright
  illumination,} Nat. Photonics \textbf{4}, 686--689 (2010).

\bibitem{wiechers2011}
C.~Wiechers, L.~Lydersen, C.~Wittmann, D.~Elser, J.~Skaar, C.~Marquardt,
  V.~Makarov, and G.~Leuchs, \enquote{After-gate attack on a quantum
  cryptosystem,} New J. Phys. \textbf{13}, 013043 (2011).

\bibitem{lydersen2010b}
L.~Lydersen, C.~Wiechers, C.~Wittmann, D.~Elser, J.~Skaar, and V.~Makarov,
  \enquote{Thermal blinding of gated detectors in quantum cryptography,} Opt.
  Express \textbf{18}, 27938--27954 (2010).

\bibitem{yuan2010}
Z.~L. Yuan, J.~F. Dynes, and A.~J. Shields, \enquote{Avoiding the blinding
  attack in {QKD},} Nat. Photonics \textbf{4}, 800--801 (2010).

\bibitem{lydersen2010c}
L.~Lydersen, C.~Wiechers, C.~Wittmann, D.~Elser, J.~Skaar, and V.~Makarov,
  \enquote{Reply to `{A}voiding the blinding attack in {QKD}',} Nat. Photonics
  \textbf{4}, 801 (2010).

\bibitem{yuan2011}
Z.~L. Yuan, J.~F. Dynes, and A.~J. Shields, \enquote{Resilience of gated
  avalanche photodiodes against bright illumination attacks in quantum
  cryptography,} Appl. Phys. Lett. \textbf{98}, 231104 (2011).

\bibitem{lydersen2011d}
L.~Lydersen, V.~Makarov, and J.~Skaar, \enquote{Comment on `{R}esilience of
  gated avalanche photodiodes against bright illumination attacks in quantum
  cryptography',} Appl. Phys. Lett. (in press); {a}rXiv:1106.3756 [quant-ph].

\bibitem{yuan2011a}
Z.~L. Yuan, J.~F. Dynes, and A.~J. Shields, \enquote{Reply to ``{C}omment on
  `{R}esilience of gated avalanche photodiodes against bright illumination
  attacks in quantum cryptography{'}{''},}  {a}rXiv:1109.3149 [quant-ph].

\bibitem{gerhardt2011}
I.~Gerhardt, Q.~Liu, A.~Lamas-Linares, J.~Skaar, C.~Kurtsiefer, and V.~Makarov,
  \enquote{Full-field implementation of a perfect eavesdropper on a quantum
  cryptography system,} Nat. Commun. \textbf{2}, 349 (2011).

\bibitem{spcmaqr}
{P}erkinElmer SPCM-AQR single photon counting module, data sheet, PerkinElmer
  (2005).

\bibitem{makarov2005}
V.~Makarov and D.~R. Hjelme, \enquote{Faked states attack on quantum
  cryptosystems,} J. Mod. Opt. \textbf{52}, 691--705 (2005).

\bibitem{bennett1984}
C.~H. Bennett and G.~Brassard, \enquote{Quantum cryptography: Public key
  distribution and coin tossing,} in \enquote{Proceedings of IEEE International
  Conference on Computers, Systems, and Signal Processing,}  (IEEE Press, New
  York, Bangalore, India, 1984), pp. 175--179.

\bibitem{rarity1994}
J.~G. Rarity, P.~C.~M. Owens, and P.~R. Tapster, \enquote{Quantum random-number
  generation and key sharing,} J. Mod. Opt. \textbf{41}, 2435--2444 (1994).

\bibitem{peloso2009}
M.~P. Peloso, I.~Gerhardt, C.~Ho, A.~Lamas-Linares, and C.~Kurtsiefer,
  \enquote{Daylight operation of a free space, entanglement-based quantum key
  distribution system,} New J. Phys. \textbf{11}, 045007 (2009).

\bibitem{hughes2002}
R.~J. Hughes, J.~E. Nordholt, D.~Derkacs, and C.~G. Peterson,
  \enquote{Practical free-space quantum key distribution over 10~km in daylight
  and at night,} New J. Phys. \textbf{4}, 43 (2002).

\bibitem{erven2008}
C.~Erven, C.~Couteau, R.~Laflamme, and G.Weihs, \enquote{Entangled quantum key
  distribution over two free-space optical links,} Opt. Express \textbf{16},
  16840--16853 (2008).

\bibitem{cova2004}
S.~Cova, M.~Ghioni, A.~Lotito, I.~Rech, and F.~Zappa, \enquote{Evolution and
  prospects for single-photon avalanche diodes and quenching circuits,} J. Mod.
  Opt. \textbf{51}, 1267--1288 (2004).

\bibitem{ralph1999}
T.~C. Ralph, \enquote{Continuous variable quantum cryptography,} Phys. Rev. A
  \textbf{61}, 010303 (1999).

\bibitem{hillery2000}
M.~Hillery, \enquote{Quantum cryptography with squeezed states,} Phys. Rev. A
  \textbf{61}, 022309 (2000).

\bibitem{reid2000}
M.~D. Reid, \enquote{Quantum cryptography with a predetermined key, using
  continuous-variable {E}instein-{P}odolsky-{R}osen correlations,} Phys. Rev. A
  \textbf{62}, 062308 (2000).

\bibitem{heid2007}
M.~Heid and N.~L\"utkenhaus, \enquote{Security of coherent-state quantum
  cryptography in the presence of {G}aussian noise,} Phys. Rev. A \textbf{76},
  022313 (2007).

\bibitem{fossier2009}
S.~Fossier, E.~Diamanti, T.~Debuisschert, A.~Villing, R.~Tualle-Brouri, and
  P.~Grangier, \enquote{Field test of a continuous-variable quantum key
  distribution prototype,} New J. Phys. \textbf{11}, 045023 (2009).

\bibitem{gol'tsman2001}
G.~N. Gol'tsman, O.~Okunev, G.~Chulkova, A.~Lipatov, A.~Semenov, K.~Smirnov,
  B.~Voronov, A.~Dzardanov, C.~Williams, and R.~Sobolewski, \enquote{Picosecond
  superconducting single-photon optical detector,} Appl. Phys. Lett.
  \textbf{79}, 705--707 (2001).

\bibitem{verevkin2002}
A.~Verevkin, J.~Zhang, R.~Sobolewski, A.~Lipatov, O.~Okunev, G.~Chulkova,
  A.~Korneev, K.~Smirnov, G.~N. Gol'tsman, and A.~Semenov, \enquote{Detection
  efficiency of large-active-area {NbN} single-photon superconducting detectors
  in the ultraviolet to near-infrared range,} Appl. Phys. Lett. \textbf{80},
  4687--4689 (2002).

\bibitem{makarov2009}
V.~Makarov, \enquote{Controlling passively quenched single photon detectors by
  bright light,} New J. Phys. \textbf{11}, 065003 (2009).

\bibitem{thew2009}
R.~T. Thew, H.~Zbinden, and N.~Gisin, \enquote{Tunable upconversion photon
  detector,} Appl. Phys. Lett. \textbf{93}, 071104 (2008).

\bibitem{thew2007}
R.~T. Thew, D.~Stucki, J.-D. Gautier, H.~Zbinden, and A.~Rochas,
  \enquote{Free-running {InGaAs/InP} avalanche photodiode with active quenching
  for single photon counting at telecom wavelengths,} Appl. Phys. Lett.
  \textbf{91}, 201114 (2007).

\bibitem{id210}
{id210} advanced system for single photon detection, data sheet, ID~Quantique
  (2011), \url{http://www.idquantique.com/images/stories/PDF/id210-single-photon-counter/id210-specs.pdf}
  (accessed on 1 August 2011).

\bibitem{dautet1993}
H.~Dautet, P.~Deschamps, B.~Dion, A.~D. MacGregor, D.~MacSween, R.~J. McIntyre,
  C.~Trottier, and P.~P. Webb, \enquote{Photon counting techniques with silicon
  avalanche photodiodes,} Appl. Opt. \textbf{32}, 3894--3900 (1993).

\bibitem{4spcmaqc}
{P}erkinElmer SPCM-AQ4C single photon counting module array, data sheet,
  PerkinElmer (2005).

\bibitem{gordon2004}
K.~J. Gordon, V.~Fernandez, P.~D. Townsend, and G.~S. Buller, \enquote{A short
  wavelength gigahertz clocked fiber-optic quantum key distribution system,}
  IEEE J. Quantum Electron. \textbf{40}, 900--908 (2004).

\bibitem{lydersen2011a}
L.~Lydersen, V.~Makarov, and J.~Skaar, \enquote{Secure gated detection scheme
  for quantum cryptography,} Phys. Rev. A \textbf{83}, 032306 (2011).

\bibitem{lydersen2011b}
L.~Lydersen, N.~Jain, C.~Wittmann, {\O}.~Mar{\o}y, J.~Skaar, C.~Marquardt,
  V.~Makarov, and G.~Leuchs, \enquote{Superlinear threshold detectors in
  quantum cryptography,} Phys. Rev. A \textbf{84}, 032320 (2011).

\bibitem{lo2011}
H.-K. Lo, M.~Curty, and B.~Qi, \enquote{Measurement device independent quantum
  key distribution,}  {a}rXiv:1109.1473 [quant-ph].

\bibitem{braunstein2011}
S.~L. Braunstein and S.~Pirandola, \enquote{Side-channel free quantum key
  distribution,}  {a}rXiv:1109.2330 [quant-ph].

\bibitem{lydersen2011c}
L.~Lydersen, M.~K. Akhlaghi, A.~H. Majedi, J.~Skaar, and V.~Makarov,
  \enquote{Controlling a superconducting nanowire single-photon detector using
  tailored bright illumination,}  {N}ew {J}. {P}hys. (in press);
  {a}rXiv:1106.2396 [quant-ph].

\end{thebibliography}
\end{document}